\documentclass[epj]{svjour}
\usepackage{graphics}
\usepackage{axodraw}

\newcommand{\calA}{{\cal A}}
\newcommand{\calU}{{\cal U}}
\newcommand{\tr}{\textrm{tr}}
\begin{document}
\title{Effects of unparticles on running of gauge couplings}
\author{Yi Liao%
\thanks{\emph{Email address:} liaoy@nankai.edu.cn}%
}                     
%
%
\institute{Department of Physics, Nankai University, Tianjin
300071, China}
\date{Received: date / Revised version: date}
%
\abstract{Unparticles charged under a gauge group can contribute
to the running of the gauge coupling. We show that a scalar
unparticle of scaling dimension $d$ contributes to the $\beta$
function a term that is $(2-d)$ times that from a scalar particle
in the same representation. This result has important implications
on asymptotic freedom. An unparticle with $d>2$, in contrast to
its matter counterpart, can speed up the approach to asymptotic
freedom for a non-Abelian gauge theory and has the tendency to
make an Abelian theory also asymptotically free. For not spoiling
the excellent agreement of the standard model (SM) with precision
tests, the infrared cut-off, $m$, of such an unparticle would be
high but might still be reachable at colliders such as LHC and
ILC. Furthermore, if the unparticle scale $\Lambda_\calU$ is high
enough, unparticles could significantly modify the unification
pattern of the SM gauge couplings. For instance, with 3 scalar
unparticles of $d\sim 2.5$ in the adjoint representation of the
strong gauge group but neutral under the electroweak one, the
three gauge couplings would unify at a scale of $\sim 8\times
10^{12}~{\rm GeV}$, which is several orders of magnitude below the
supersymmetric unification scale.
} 
\maketitle

Asymptotic freedom \cite{Gross:1973id} had been historically
stimulated by the suggestion \cite{Bjorken:1968dy} and observation
\cite{Bloom:1969kc} of Bjorken scaling in deeply inelastic
scattering, and contributed significantly to the establishment of
quantum chromodynamics, a non-Abelian gauge theory, as the correct
theory of strong interactions. It further offered the insight to
the endeavors \cite{Georgi:1974sy} to unify strong and electroweak
interactions: These interactions, though very different in their
strength at low energies, may originate from a unified theory with
a single coupling constant at some high energy scale
\cite{Georgi:1974yf} since the strength of strong interactions
decreases as energy increases.

Central to the idea of asymptotic freedom is the negative sign of
the $\beta$ function that determines renormalization group running
of a coupling. It has been well established in quantum field
theory \cite{Coleman:1973sx} that only non-Abelian gauge theory
can have a negative $\beta$ if it does not contain too many matter
fields. The latter restriction is necessary since all matter
fields contribute positively to the $\beta$ function. In
particular, an Abelian gauge theory that can only interact with
the matter fields are not asymptotically free.

Nevertheless, we want to show in this work that something very
different from the stated can occur when gauge bosons are coupled
to some effective degrees of freedom arising from certain scale
invariant sector at higher energies. Unparticles as suggested
recently by Georgi \cite{Georgi:2007ek} provide a concrete
realization of such a degree of freedom, on which our work is
based. We find that an unparticle charged under a gauge group
contributes to the $\beta$ function of the gauge coupling a term
that is a multiple of the one from a matter field charged
identically under the group. The multiplication factor can be of
either sign, depending on the scaling dimension of the unparticle.
This result has important impact on asymptotic freedom. An
unparticle with an appropriate scaling dimension can speed up the
approach to asymptotic freedom of a non-Abelian theory, and can
make an Abelian theory become asymptotically free. If unparticles
are charged under the standard model (SM) gauge group, they will
modify significantly the pattern of the grand unification.

Although our finding is surprising and in sharp contrast to the
conventional statement on matter fields, we do not see any obvious
conflict between the two. The reason is that the unparticle field
is not a field in the conventional sense. Its quantum, unparticle,
is totally different from the particle quantum of a conventional
field. It does not enjoy mass as one of its defining properties;
instead, its kinematics is largely determined by the scaling
dimension of its field, $d$, a generally non-integral number. The
latter makes the unparticle field a non-local object that
interacts differently from a conventional local field.

The original work of Ref \cite{Georgi:2007ek} has triggered
intensive activities in unparticle physics in the past few months.
Many of its salient features have been unveiled, investigated and
applied \cite{Georgi:2007si}-\cite{Bhattacharyya:2007pi} through
interactions with the SM particles. Most of these interactions
imply implicitly that unparticles are actually charged under the
SM gauge group, $G_{SM}=SU(3)_c\times SU(2)_L\times U(1)_Y$, but
no attempt has been made to incorporate direct gauge interactions
of unparticles until very recently in Ref
\cite{Cacciapaglia:2007jq}. Although this issue has already been
challenged in \cite{Georgi:2007ek}, it is non-trivial because of
the non-local nature of the unparticle field. Fortunately, a
similar problem was successfully treated some years ago in
modeling low energy chiral dynamics of Goldstone bosons by
dynamical quarks \cite{Terning:1991yt,Holdom:1992fn}. A dynamical
quark with a non-trivial momentum-dependent effective mass implies
a non-local term in its Lagrangian which, when gauged, yields
non-local interactions between the dynamical quark and gauge
fields. The first phenomenological, interesting result has been
obtained in \cite{Cacciapaglia:2007jq}. We shall demonstrate that
this non-local gauging of unparticle fields has far-reaching
implications on the running properties of the gauge couplings
themselves.

The scale invariance of the unparticle field determines the
density of states in phase space of a scalar unparticle of
momentum $p$ to be proportional to
$\theta(p^0)\theta(p^2)(p^2)^{d-2}$. Then unitarity considerations
imply the following propagator \cite{Georgi:2007si,Cheung:2007ue}:
\begin{eqnarray}
iD(p)=\frac{A_d}{2\sin(\pi d)}\frac{i}{(-p^2-i\epsilon)^{2-d}},
\end{eqnarray}
where $A_d$ is a normalization factor inessential to our
discussion. If scale invariance does not extend to arbitrarily low
energy, some modifications to the above are required. The simplest
way would be to add an infrared cut-off $m^2$ to the inside of the
above power \cite{Fox:2007sy}, and we shall follow this ansatz
although this will not affect our result in the ultraviolet.

The propagator is supposed to be derived from a Lagrangian
quadratic in the unparticle field $\calU(x)$. The latter is
generally non-local for $d$ non-integral or greater than $2$, with
the action being
\begin{eqnarray}
S_0=\int d^4x~d^4y~\calU^{\dagger}(x)\tilde{D}^{-1}(x-y)\calU(y),
\end{eqnarray}
where $\tilde{D}^{-1}(z)$ is the Fourier transform of $D^{-1}(p)$.
When $\calU$ is charged under a local gauge group, the usual
minimal coupling for a local field does not yield a gauge
invariant term. Instead, a convenient way to build an invariant
form is to invoke the Wilson line \cite{Cacciapaglia:2007jq}:
\begin{eqnarray}
S&=&\int d^4x~d^4y~\calU^{\dagger}(x)\tilde{D}^{-1}(x-y)
\nonumber\\
&\times&P\exp\left[-igT^a\int_x^yA_{\mu}^a~dw^{\mu}\right]\calU(y),
\end{eqnarray}
where $P$ denotes path-ordering in the generators $T^a$ of the
gauge group in the unparticle representation. A systematic
approach was developed in \cite{Terning:1991yt,Holdom:1992fn} to
derive interaction vertices from such an action.

For our purpose of computing the unparticle contribution to the
$\beta$ function, we consider the vacuum polarization diagrams
shown in Fig. 1, which require the following two vertices:
\begin{eqnarray}
&&ig\Gamma^{a\mu}(-(p+q),p;q)\nonumber\\
&=&igT^a(2p+q)^{\mu}E_1(p;q),\nonumber\\
&&ig^2\Gamma^{ab\mu\nu}(-(p+q_1+q_2),p;q_1,q_2)\nonumber\\
&=&ig^2\left\{
\{T^a,T^b\}g^{\mu\nu}E_1(p;q_1+q_2)\right.\nonumber\\
&+&T^aT^b(2p+q_2)^{\nu}(2p+2q_2+q_1)^{\mu}E_2(p;q_2,q_1)\nonumber\\
&+&\left.T^bT^a(2p+q_1)^{\mu}(2p+2q_1+q_2)^{\nu}E_2(p;q_1,q_2)\right\},
\end{eqnarray}
where all momenta are meant to be incoming with the first two for
the unparticles and others for the gauge bosons, and the recursive
form factors are
\begin{eqnarray}
E_0(p)&=&D^{-1}(p),\nonumber\\
E_1(p;q_1)&=&\frac{E_0(p+q_1)-E_0(p)}{(p+q_1)^2-p^2},\nonumber\\
E_2(p;q_1,q_2)&=&\frac{E_1(p;q_1+q_2)-E_1(p;q_1)}{(p+q_1+q_2)^2-(p+q_1)^2}.
\end{eqnarray}
The case of scalar particles is nicely recovered in the limit
$d\to 1$ by noting that $\frac{A_d}{2\sin(\pi d)}\to -1,~E_1\to 1$
and that the recursion terminates at $E_2\to 0$.
\begin{center}
\begin{picture}(200,160)(0,0)

\SetOffset(50,90)
\DashCArc(40,40)(25,0,360){3}\DashCArc(40,40)(24,0,360){2.5}
\Photon(0,40)(15,40){2}{3}\Photon(65,40)(80,40){2}{3}%
\Vertex(15,40){2.5}\Vertex(65,40){2.5}
\Text(82,40)[l]{{\footnotesize$-q,\nu,b$}}%
\Text(-2,40)[r]{{\footnotesize$q,\mu,a$}}%
\Text(18,20)[r]{{\footnotesize$p$}}%
\Text(40,0)[]{$(1)$}

\SetOffset(50,10) %
\DashCArc(40,40)(25,0,360){3}\DashCArc(40,40)(24,0,360){2.5}
\Photon(0,15)(80,15){2}{14}\Vertex(40,15){2.5}%
\Text(40,0)[]{$(2)$}

\end{picture}

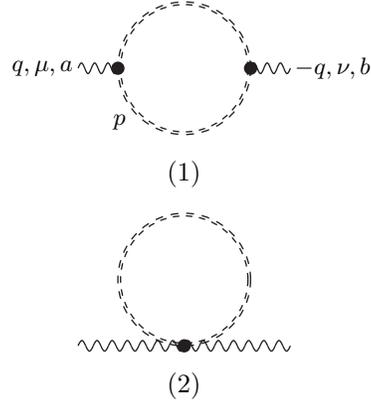
{\small Fig. 1. Unparticles contributions to the vacuum
polarization.}
\end{center}

The imaginary part of the diagrams has been computed in
\cite{Cacciapaglia:2007jq} in a sophisticated manner. We shall do
a complete calculation for the whole diagrams, and the result
turns out to be surprisingly simple. The first diagram is
\begin{eqnarray}
i\calA_1^{\mu\nu}&=&g^2\tr T^aT^b
\int\frac{d^np}{(2\pi)^n}\frac{(2p+q)^{\mu}(2p+q)^{\nu}}{(2p\cdot
q+q^2)^2}\nonumber\\
&\times&
\left[\frac{D(p)}{D(p+q)}+\frac{D(p+q)}{D(p)}-2\right],
\end{eqnarray}
which is symmetric in $(2-d)\leftrightarrow(d-2)$ as can be seen
by $p\to -(p+q)$. We work in $n$ dimensions, where the power in
$D(p)$ should be replaced by $\frac{n}{2}-d$ for consistency, but
this does not affect the extraction of the $\beta$ function. The
second diagram is
\begin{eqnarray}
i\calA_2^{\mu\nu}&=&-g^2\tr T^aT^b\int\frac{d^np}{(2\pi)^n}D(p)
\left\{2g^{\mu\nu}E_1(p;0)\right.\nonumber\\
&&+(2p-q)^{\nu}(2p-q)^{\mu}E_2(p;-q,q)\nonumber\\
&&+\left.(2p+q)^{\mu}(2p+q)^{\nu}E_2(p;q,-q)\right\},
\end{eqnarray}
where
\begin{eqnarray}
E_1(p;0)&=&-\frac{2-d}{m^2-p^2-i\epsilon}\frac{1}{D(p)}.
\end{eqnarray}
The second term in $i\calA_2^{\mu\nu}$ can be made equal to the
third by $p\to -p$ and using $E_2(-p;-q,q)=E_2(p;q,-q)$. A little
algebra gives
\begin{eqnarray}
i\calA_2^{\mu\nu}&=&-2g^2\tr T^aT^b\int\frac{d^np}{(2\pi)^n}
\nonumber\\
&\times&\left\{\left[\frac{(2p+q)^{\mu}(2p+q)^{\nu}}{2q\cdot
p+q^2}-g^{\mu\nu}\right]\frac{2-d}{m^2-p^2-i\epsilon}\right.\nonumber\\%
&&\left.+\frac{(2p+q)^{\mu}(2p+q)^{\nu}}{(2q\cdot p+q^2)^2}
\left[\frac{D(p)}{D(p+q)}-1\right]\right\},
\end{eqnarray}
where the first term is odd in $(2-d)$. By $p\to-(p+q)$, the
second can be cast in a form that amounts to $(2-d)\to(d-2)$, and
averaging it with the original form yields exactly
$-i\calA_1^{\mu\nu}$. Thus we have
\begin{eqnarray}
i(\calA_1+\calA_2)^{\mu\nu}&=&-2g^2\tr
T^aT^b\int\frac{d^np}{(2\pi)^n}
\frac{2-d}{m^2-p^2-i\epsilon}\nonumber\\
&\times&\left[\frac{(2p+q)^{\mu}(2p+q)^{\nu}}{2q\cdot
p+q^2}-g^{\mu\nu}\right].
\end{eqnarray}
Applying again the same trick to the first term in the above sum
yields the final answer:
\begin{eqnarray}
&&i(\calA_1+\calA_2)^{\mu\nu}\nonumber\\
&=&(2-d)g^2\tr T^aT^b
\left[\int\frac{d^np}{(2\pi)^n}\frac{2g^{\mu\nu}}{m^2-p^2-i\epsilon}
\right.\nonumber\\
&&\left.+\int\frac{d^np}{(2\pi)^n}\frac{(2p+q)^{\mu}(2p+q)^{\nu}}
{[m^2-(p+q)^2-i\epsilon][m^2-p^2-i\epsilon]}\right].
\end{eqnarray}
Namely, the scalar unparticle contribution to the vacuum
polarization is $(2-d)$ times that of the scalar particle in the
same representation.

A few comments are in order. The above manipulation is much
simpler than that in \cite{Cacciapaglia:2007jq} while yielding an
even stronger result: the relation between the unparticle and
particle contributions holds only just for the imaginary part as
explicitly shown there, but for the whole amplitude. This result
is consistent with the `slick' argument in that paper based on
path integrals which however could be dangerous due to the
presence of ultraviolet singularities and non-locality. The
relation for the imaginary part and the conventional optical
theorem for particles were further employed in
\cite{Cacciapaglia:2007jq} to obtain the cross section for pair
production of colored unparticles via one gluon exchange from the
initial quark and anti-quark state. A potential problem with the
naive application of the theorem in the present model is discussed
in the Appendix. The conclusion drawn from the discussion is that
the subtlety does not pose an obstacle to our main interest of
calculating unparticle effects on the running of gauge couplings.

Having obtained the vacuum polarization, the contribution to the
$\beta$ function can be written down directly for $n_{\calU}$
species of unparticles in the representation $r_{\calU}$ of the
gauge group:
\begin{eqnarray}
\beta(g)_{\calU}=(2-d)\frac{1}{4}n_{\calU}\cdot
\frac{g^3}{(4\pi)^2}\frac{4}{3}C(r_{\calU})
\end{eqnarray}
where $(2-d)$ is as computed, $\frac{1}{4}$ for scalars and the
last factors are standard for a fermion particle multiplet in the
representation $r_{\calU}$ with $\tr
T^aT^b=C(r_{\calU})\delta^{ab}$. This result is simple and
surprising. As argued in Ref \cite{Georgi:2007ek}, an unparticle
with scale dimension $d$ kinematically looks like a number $d$ of
invisible massless particles, but its contribution to the $\beta$
function does not look like a matter field but something opposite:
the $d$ term has a minus sign. That $\beta(g)_{\calU}$ happens to
vanish at $d=2$ is a result in four dimensions; at this value all
interactions are away as is clear from the action. Concerning $d$,
there are no real constraints but the one from analysis of unitary
representations in conformal theory: $d\ge 1$ for a scalar object
\cite{Mack:1975je}. Thus, the unparticle term could be of either
sign.

Particularly interesting is the possibility that $d>2$. In this
case, unparticles behave oppositely to matter fields in affecting
the running of a gauge coupling. A non-Abelian gauge theory can
approach the asymptotic freedom faster when it is coupled to such
unparticles fields. An Abelian gauge theory, which is otherwise
not asymptotically free, could become asymptotically free if the
unparticle contribution dominates over that of matter fields. At
first sight, this might look like a mere academic interest since
the running of the SM gauge couplings has been well tested, see
for instance Ref \cite{Bethke:2006ac} for a recent review on
experimental tests of asymptotic freedom in QCD. This is not
necessarily true. Unparticles as a remnant of some scale invariant
theory at high energy could become relevant at a scale that is a
bit higher than a few hundred GeV, up to which the running of the
SM gauge couplings has only been measured. To avoid spoiling the
well-tested region, we thus require an infrared cut-off $m$ for
unparticles that is high enough. Supposing that the unparticle
scale $\Lambda_\calU$ is much higher than the electroweak scale,
we shall thus investigate how the unification of the gauge
couplings in SM could be affected at high energy scales in
between.

The $\beta$ functions for the three gauge couplings in SM are
\begin{eqnarray}
&&\beta(g_s)_{\rm SM}=-\frac{g_s^3}{4\pi^2}\frac{7}{4},~
\beta(g)_{\rm
SM}=-\frac{g^3}{4\pi^2}\frac{5}{6},\nonumber\\
&&\beta(g')_{\rm SM}=\frac{g^{\prime 3}}{4\pi^2}\frac{5}{3}
\end{eqnarray}
The unparticle terms should be added to the above accordingly when
they are charged under a gauge group. At the scale $M$, those
couplings are expected to unify
\cite{Georgi:1974sy,Georgi:1974yf}:
\begin{eqnarray}
g_s^2(M)=g^2(M)=\frac{5}{3}g^{\prime 2}(M)
\end{eqnarray}
Their values at the $Z$ boson mass have been well determined. For
those, we use the numbers from Particle Data Group:
$g^{-2}_s(m_Z)=0.654$, $g^{-2}(m_Z)=2.354$, $g^{'-2}(m_Z)=7.826$
with $m_Z=91.19~{\rm GeV}$. For illustration purpose, we consider
a simple scenario that contains $n_{\calU}$ species of scalar
unparticles with dimension $d$, all in the same representation
$r_{\calU}$ of $SU(3)_c$ but neutral under $SU(2)_L\times U(1)_Y$.
In other words, $g^{-2}$ and $\frac{5}{3}g^{\prime-2}$ meet at the
same scale as would in SM, $M\sim 8\times 10^{12}~{\rm GeV}$,
where $g^{-2}(M)=3.418$.
\begin{center}
\begin{picture}(250,180)(0,0)
\SetOffset(70,45) %
\SetScale{0.4}%
\LinAxis(0,0)(400,0)(4,10,2,0,1)\LinAxis(0,300)(400,300)(4,10,-2,0,1)%
\LinAxis(0,0)(0,300)(6,5,-2,0,1)\LinAxis(400,0)(400,300)(6,5,2,0,1)%
\DashLine(0,191.8)(400,191.8){9}
\DashCurve{(0,128)(50,131)(100,134)(150,136)(200,139)(250,142)(300,144)(350,147)(400,150)}{1}
\DashCurve{(0,107)(50,115)(100,123)(150,131)(200,139)(250,147)(300,155)(350,163)(400,171)}{3}
\DashCurve{(0,85)(50,99)(100,112)(150,126)(200,139)(250,152)(300,166)(350,179)(400,192)}{5}
\Curve{(0,75)(50,91)(100,107)(150,123)(200,139)(250,155)(300,171)(350,187)(400,203)}
\Curve{(55,0)(100,43)(150,91)(200,139)(250,187)(300,235)(350,282)(368,300)}
\Text(0,-7)[]{0}\Text(40,-7)[]{1}\Text(80,-7)[]{2}%
\Text(120,-7)[]{3}\Text(160,-7)[]{4}\Text(80,-20)[]{$d$}%
\Text(-7,0)[r]{1.5}\Text(-7,20)[r]{2.0}\Text(-7,40)[r]{2.5}\Text(-7,60)[r]{3.0}
\Text(-7,80)[r]{3.5}\Text(-7,100)[r]{4.0}\Text(-7,120)[r]{4.5}
\Text(-25,60)[r]{$g^{-2}_s(M)$}%
\Text(5,115)[l]{{\footnotesize$(r_{\calU},n_{\calU})$}}
\Text(3,55)[l]{{\tiny$({\bf 3},1)$}}%
\Text(3,48)[l]{{\tiny$({\bf 3},3)$}}%
\Text(3,40)[l]{{\tiny$({\bf 3},5)$}}%
\Text(55,42)[]{{\tiny$(G,1)$}}%
\Text(55,20)[]{{\tiny$(G,3)$}}
\end{picture}

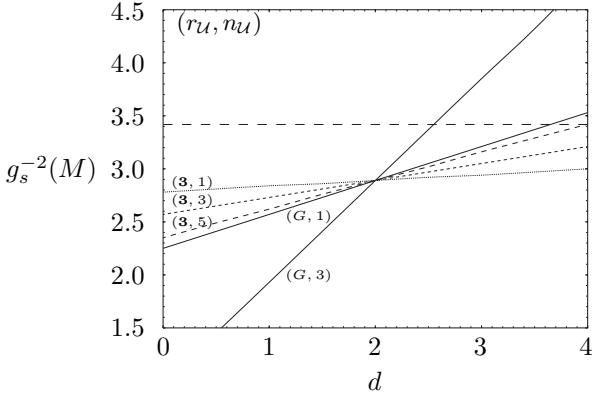
{\small Fig. 2: $g^{-2}_s(M)$ as a function of $d$ for $n_{\calU}$
species of scalar unparticles in the fundamental or adjoint
representation of $SU(3)_c$.}
\end{center}

The value of $g^{-2}_s(M)$ at the scale $M$ is shown in Fig. 2 as
a function of $d$ for $n_{\calU}$ species of scalar unparticles in
the fundamental (${\bf 3}$) or adjoint ($G$) representation of
$SU(3)_c$. The horizontal dash line corresponds to the point where
all three gauge couplings unify with the help of these
unparticles. It is interesting that with a few simply arranged
unparticles the unification scale for the SM gauge couplings could
be several orders of magnitude lower than the one in
supersymmetric standard model \cite{Langacker:1991an}. Unparticle
physics may provide an alternative to ordinary new physics for
grand unification.

Unparticles should interact with particles observed in experiments
to be physically relevant. It is natural they carry charges under
the SM gauge group. Due to non-trivial scale dimensions of
unparticle fields, their gauge theory is generally non-local.
While the interactions in such a theory may be complicated, some
properties are reachable without ambiguities. Although these
interactions may back-react to break scale invariance of
unparticles below the electroweak scale or due to renormalization
effects, their leading effects to the running of gauge couplings
from the electroweak scale to the unparticle scale can be
consistently studied. We have shown that an unparticle field
contributes to the $\beta$ function of a gauge coupling in a
simple manner: the ordinary result for a matter field multiplied
by a factor depending on the unparticle's scale dimension. The
point is that this factor could be of either sign, instead of
being positive-definite as in the matter case. An Abelian gauge
theory could become asymptotically free if the gauge boson
interacts with an unparticle of a large enough scale dimension.
This new feature in the $\beta$ function could have implications
in the other context of gauge theory.

The effect on the $\beta$ function could potentially modify the
pattern of unification of gauge couplings. We have illustrated
this with a simple scenario showing that the unification scale
could be made several orders of magnitude lower than the one in
supersymmetric unification. But there are many uncertain factors
with this unparticle-assisted unification. It depends on the
detailed arrangement of unparticles under the SM gauge group, to
which there currently seems to be no guide. More uncertain is
perhaps the impact from high energy physics that produces
unparticles; in our illustrative example, we have implicitly
assumed that the unification scale is lower than the scale at
which unparticle degrees of freedom appear.

There are several things worthy to be explored. Unparticle effects
could be avoided by setting a large enough cut-off $m$. But
unparticles will become more relevant and interesting if it is not
too high. For instance, it might be possible that some effects are
observable at high energy colliders such as the LHC or ILC without
spoiling the precision data at lower energies. They could also be
detected in ultra-high energy processes in astrophysics. All of
this will depend on a more or less complete model for unparticles
in the framework of electroweak and strong interactions.

\noindent %
{\bf Acknowledgement} I would like to thank Prof. Xiaoyuan Li for
useful discussions. This work was supported in part by the grants
NCET-06-0211 and NSFC-10775074.

\noindent %
{\bf Appendix}

We show by symmetry analysis that the optical theorem for particle
scattering very likely breaks down for unparticles in the present
model when it is straightforwardly applied. We take the same
process computed in \cite{Cacciapaglia:2007jq}, i.e.,
$q(p_1)\bar{q}(p_2)\to\calU(k_1)\bar{\calU}(k_2)$, where quarks
are treated massless and the unparticle $\calU$ has the color
representation $r_{\calU}$ with the generators $T^a$ normalized as
$\tr T^aT^b=C(r_{\calU})\delta^{ab}$. The amplitude is
\begin{eqnarray}
i\calA&=&\bar{v}(p_2)ig_s\gamma_{\mu}\frac{\lambda^a}{2}u(p_1)\frac{-i}{(p_1+p_2)^2}
ig_sT^a\frac{(k_1-k_2)^{\mu}}{k_1^2-k_2^2}\nonumber\\
&\times&\left[\frac{1}{D(k_1)}-\frac{1}{D(k_2)}\right].
\end{eqnarray}
Note that $D(k_i)$ is complex for $K_i=k_i^2-m^2>0$:
\begin{eqnarray}
D(k_i)&=&\frac{A_d}{2\sin(d\pi)}K_i^{d-2}e^{-i(\pi-\epsilon)d}.
\end{eqnarray}
Doing color and spin summation and averaging and attaching the
phase space factors for unparticles, the differential cross
section is
\begin{eqnarray}
d\sigma&=&\sin^2(d\pi)(N_c^2-1)C(r_{\calU})\frac{g_s^4}{9s^3}\nonumber\\
&\times&
\frac{d^4k_1}{(2\pi)^4}\frac{d^4k_2}{(2\pi)^4}(2\pi)^4\delta^4(p_1+p_2-k_1-k_2)
\nonumber\\
&\times&\left[2p_1\cdot(k_1-k_2)p_2\cdot(k_1-k_2)-p_1\cdot
p_2(k_1-k_2)^2\right]\nonumber\\
&\times&\frac{1}{(K_1-K_2)^2}\left[\left(\frac{K_1}{K_2}\right)^{2-d}
+\left(\frac{K_2}{K_1}\right)^{2-d}-2\right],
\end{eqnarray}
where the step functions for $k_i$ are implied and $N_c=3$. The
apparent singularity at $k_1^2=k_2^2$ can be removed using a trick
\cite{Cacciapaglia:2007jq}:
\begin{eqnarray}
&&\left(\frac{K_1}{K_2}\right)^{2-d}=
K_1\frac{1}{K_1^{d-1}K_2^{d-2}}\nonumber\\
&=&\frac{K_1}{\Gamma(d-1)\Gamma(2-d)}\int_0^1dx
\frac{x^{1-d}(1-x)^{d-2}}{K_1(1-x)+K_2x},
\end{eqnarray}
so that all $K_i$ factors combine to
\begin{eqnarray}
&&\frac{1}{\Gamma(d-1)\Gamma(2-d)}\nonumber\\
&\times&\int_0^1dx
\frac{x^{2-d}(1-x)^{d-2}(2x-1)}{[K_1(1-x)+K_2x][K_2(1-x)+K_1x]}.
\end{eqnarray}
For symmetry analysis, we keep only relevant factors:
\begin{eqnarray}
d\sigma&\propto&\sin^3(d\pi)\nonumber\\
&\times&\int_0^1dx
\frac{x^{2-d}(1-x)^{d-2}(2x-1)}{[K_1(1-x)+K_2x][K_2(1-x)+K_1x]}.
\end{eqnarray}
The pre-factor changes sign when $(2-d)\to(d-2)$, so does the
integral as can be seen by $x\to(1-x)$. The cross section thus
obtained is an even function of $(2-d)$, in contrast to the vacuum
polarization which is odd. Its first non-vanishing derivative at
$d=2$ appears in the fourth order, and thus the cross section is
not a linear function of $d$ in the neighborhood of $d=2$. This
would be sufficient to signal the breakdown of the particle
optical theorem in the unparticle gauge model.

This by itself should not be too surprising. Unparticles do not
correspond to asymptotic states in the usual sense of the word,
for which the issue of how to define an $S$ matrix is not yet
settled. The calculations in the literature on processes involving
unparticles in the initial or final state seem to go through the
check of the naive optical theorem, as the author has checked, but
all those are restricted to unparticles that are not charged under
any gauge group. The new element in the gauge model of
\cite{Cacciapaglia:2007jq} is the presence of the propagator
$D(p)$ in interaction vertices. For non-integral $d$, this
modifies the analyticity properties in a non-trivial way. For
instance, the interaction vertices are even not Hermitian when an
involved unparticle is above its infrared cut-off, $p^2>m^2$.

The situation is similar to that in the non-local chiral quark
model. One of the purposes to introduce dynamical quark mass had
been to mimic the QCD effects on chiral dynamics of Goldstone
bosons and to compute the low energy constants in chiral
Lagrangian in particular \cite{Holdom:1990iq}. A form for
dynamical quark mass is generally assumed for Euclidean momentum,
whose limit in the deep Euclidean region is well motivated. This
is sufficient for dynamics of Goldstone bosons, and no problem is
anticipated when dynamical quarks are confined in loops. This is
also the case here: the unparticle is treated as an interpolating
field or effective degree of freedom from certain scale invariant
physics whose loop effects on dynamics of gauge fields are
studied. No problem is thus expected for the $\beta$ function
calculated in this paper that is related to the real part of the
vacuum polarization due to virtual unparticles. When trying to
compute quantities of a `free' quark, however, e.g., the axial
vector coupling of the constituent quark \cite{Li:1993et}, care
must be taken and additional assumptions are required for
dynamical quark mass in the time-like region.

The possible breakdown of the optical theorem due to the change of
analyticity properties has a well-known analog in non-commutative
field theory. When Feynman rules are naively derived, the optical
theorem is indeed broken when time does not commute with space
\cite{Gomis:2000zz}. The breakdown can be attributed to the
appearance of phase factors that involve the zero component of
momentum in the naive Feynman rules and modify analyticity
properties of Green functions in a significant way
\cite{Liao:2002xc}. A careful treatment with them results in
Feynman rules fulfilling the optical theorem.

%


\end{document}